\documentclass[a4paper,english,aps]{revtex4}
\usepackage[T1]{fontenc}
\usepackage{amsmath}
\usepackage[dvips]{graphicx}
\usepackage{amssymb}

\makeatletter

\providecommand{\tabularnewline}{\\}

\usepackage{babel}
\makeatother
\begin{document}

\title{Energy Transfer and Bottleneck Effect in Turbulence}

\author{Mahendra K. Verma}

\affiliation{Department of Physics, Indian Institute of Technology, Kanpur 208016,
India}

\author{Diego Donzis}

\affiliation{Georgia Institute of Technology, School of Aerospace Engineering,
Atlanta, GA 30332, USA}

\begin{abstract}
Past numerical simulations and experiments of turbulence exhibit a
hump in the inertial range, called the bottleneck effect. In this
paper we show that sufficiently large inertial range (four decades)
is required for an effective energy cascade. We propose that the bottleneck
effect is due to the insufficient inertial range available in the
reported simulations and experiments. To facilitate the turbulent
energy transfer, the spectrum near Kolmogorov's dissipation wavenumber
has a hump.
\end{abstract}
\maketitle

\section{Introduction}

Energy spectrum of turbulent flow is an important quantity. In 1941,
Kolmogorov \cite{K41a} showed that the energy spectrum $E(k)$ of
turbulent flow is \begin{equation}
E(k)=K_{Ko}\Pi^{2/3}k^{-5/3}f(k/k_{d}),\label{eq:Kolm-Ek}\end{equation}
where $\Pi$ is the energy flux, $K_{Ko}$ is Kolmogorov's constant,
$k_{d}$ is Kolmogorov's wavenumber, and the function $f(x)\rightarrow1$
in the inertial range ($x\ll1$), and $f(x)\rightarrow0$ as $x\gg1$.
Many experiments and numerical simulations verify this powerlaw apart
from a very small intermittency correction. The compensated energy
spectrum $E(k)k^{5/3}/K_{Ko}$ is flat in the inertial range, and
decays in the dissipation range. A careful observation of energy spectrum
obtained from recent high-resolution numerical simulations and experiments
however show a small hump near Kolmogorov's wavenumber $k_{d}$. The
feature is called the \emph{bottleneck effect } in literature. In
this paper we propose an explanation for the bottleneck effect.

The bottleneck effect has been reported in many numerical simulations
and experiments of fluid turbulence. Yeung and Zhou \cite{YeunZhou},
Gotoh et al. \cite{Goto:VeloStat}, Kaneda et al. \cite{Kane:Bottle},
and Dobler et al. \cite{Bran:Bottle} found the bottleneck effect
(hump in the normalized energy spectrum) in their numerical simulations.
Saddoughi and Veeravalli \cite{Veer:Bottle} studied the energy spectrum
of atmospheric turbulence and reported the bottleneck effect. They
observed that the longitudinal spectra have a larger inertial range
(around 1.5 decade) but smaller hump, while the transverse spectra
have relatively smaller inertial range (around one decade), but a
larger hump. Shen and Warhaft \cite{Shen}, Pak et al. \cite{Pak:Bottle},
She and Jackson \cite{She:Bottle}, and other experimental groups
also observed the bottleneck effect in fluid turbulence.

The bottleneck effect has been seen in other forms of turbulence as
well. Watanabe and Gotoh and others \cite{Gotoh:Scalar_Bottle,Yeung:JFM96,DonzKRS:Scalar}
reported the bottleneck effect in scalar turbulence, Haugen et al.
\cite{Bran:Flux} in three-dimensional magnetohydrodynamics, Biskamp
et al. \cite{Bisk:Bottle_2D} in two-dimensional magnetohydrodynamics,
electron-magnetohydrodynamics, and thermal convection. Lamorgese et
al. \cite{Pope:Hypervisc_Bottle}, Biskamp et al. \cite{Bisk:Bottle_2D},
and Dobler et al. \cite{Bran:Bottle} observed that the bottleneck
effect became more pronounced when hyperviscosity is increased. 

There have been various attempts to explain bottleneck effect.  Falkovich
\cite{Falk:Bottleneck} argued that the viscous suppression of small-scale
modes removes some triads from nonlinear interactions, thus making
it less effective, which leads to pileup of energy in the inertial
interval of scales. Based on turbulent viscosity and the assumption
of local energy transfer, Falkovich derived the following formula
for the correction in Kolmogorov's spectrum:\begin{equation}
\delta E(k)=E(k)(k/k_{p})^{4/3}/\ln(k_{p}/k),\label{eq:Falk}\end{equation}
where $k_{p}$ is proportional to the dissipative wavenumber $k_{d}$,
and $k\ll k_{p}$. 

Yakhot and Zakharov \cite{YakhZakh:Cleb} derived energy spectrum
using Clebsch variables and showed that the energy spectrum is\begin{equation}
E(k)=K_{Ko}\Pi^{2/3}k^{-5/3}f(k/k_{d})+Pk^{-1},\label{eq:Yakh}\end{equation}
i.e., correction is of the form $k^{-1}$. Theoretical justification
for $k^{-1}$ was argued by Orszag \cite{Orsz:Rev} who analyzed the
one-loop Dyson equation for the propagator $G$ and the velocity correlation
function $U$; the spectrum $k^{-1}$ was obtained by assuming that
the response function is dominated by viscous effects. She and Jackson
\cite{She:Bottle} reported an experimental result in which they observed
the $k^{-1}$ bottleneck correction; they argued  coherent vortex
structures to be the reason for the bottleneck effect.

Kurien et al. \cite{Kuri:Bottle} extended Kolmogorov's phenomenology
to include the effects of helicity. They found a shallower $k^{-4/3}$
energy spectrum at higher wavenumbers by assuming that the helicity
transfer time-scales dominate at large wavenumbers. In the following
discussion we will propose a new mechanism to explain the bottleneck
effect. We argue that the bottleneck effect is seen when the length
of inertial range is insufficient for the energy cascade process.

\section{The Reason for the Bottleneck Effect}

The basic idea presented in our paper is as follows: In a fully-developed
turbulence, a flux of energy is transferred from small wavenumbers
to large wavenumbers. This process involves interactions of large
number of modes---from small wavenumbers to large wavenumbers. The
maximum energy transfer from a given wavenumber shell is to its nearest
neighbour. Still significant amount of energy transfer takes place
between somewhat distant wavenumber shells \cite{Krai:71,Zhou:Local,Zhou:RevLocal,Doma:Local2,Wale,Ayye}.
Verma et al. \cite{Ayye} showed using a theoretical arguments that
if the inertial-range shells are divided in such a way that the $m$th
shell is given by $k_{0}(2^{m/4}:2^{(m+1)/4})$, then in the inertial-range,
the normalized shell-to-shell energy transfer rates from shell $m$
to shell $m+1,m+2,m+3$ are $18\%,6.7\%,$ and $3.6\%$ respectively.
The remaining portion of energy flux, which is a huge fraction ($\sim$$70\%$),
is transferred to the distant shells. This result is in agreement
with earlier simulation results \cite{Zhou:Local}. 

The above arguments imply that for an effective cascade of energy,
there must be a large enough range of wavenumbers. Ideally, when Kolmogorov's
wavenumber $k_{d}\rightarrow\infty$, Kolmogorov's cascade is setup,
and the energy spectrum is given by Eq. (\ref{eq:Kolm-Ek}). However,
if Kolmogorov's wavenumber is not sufficiently large, the cascade
process faces difficulty; at higher wavenumbers there are not enough
number of modes to receive the energy transferred from the smaller
wavenumbers. To compensate, the wavenumbers near Kolmogorov's scale
have a higher energy level. We propose this to be the main reason
for the bottleneck effect. Note that the energy fed at small wavenumbers
fixes the level of energy spectrum in the inertial range, and the
energy input has to be dissipated at the higher wavenumbers. In the
following discussion we will present a quantitative arguments to support
the above idea.

\subsection{Formalism}

The average energy flux from a wavenumber sphere of radius $k_{0}$
is given by \cite{Lesi:book,Lesl:book,MKV:PR}\begin{equation}
\Pi(k_{0})=\int_{k>k_{0}}d\mathbf{k}\int_{p<k_{0}}d\mathbf{p}\left\langle S(k|p|q)\right\rangle \end{equation}
 where $S(k|p|q)$ is the {}``\emph{mode-to-mode energy transfer
rate}'' in a triad (\textbf{p,q,k)} with $\mathbf{k}=\mathbf{p+q}$,
and $\left\langle \right\rangle $ represents the ensemble average.
The term $S(k|p|q)$ represents the energy transfer rate from mode
$\mathbf{p}$ to mode $\mathbf{k}$ with mode \textbf{q} acting as
a mediator. The term $\left\langle S(k|p|q)\right\rangle $ has been
computed earlier using standard field-theoretic technique \cite{Lesi:book,Lesl:book,McCoWatt,McCo:book,MKV:PR}.
The procedure to compute $\Pi(k_{0})$ is described in the above references,
which yields\begin{equation}
\Pi(k_{0})=K_{Ko}^{3/2}\Pi\left[\int_{k_{0}}^{\infty}dkk{}^{2}\int_{0}^{k_{0}}dp\int_{|k-p|}^{k+p}dq\frac{{pq}}{4k}\frac{{T_{1}C(p)C(q)+T_{2}C(q)C(k)+T_{3}C(p)C(k)}}{\nu^{*}(k^{2/3}+p^{2/3}+q^{2/3})}\right],\label{eq:Pik0}\end{equation}
where $\nu^{*}$ is the renormalized parameter in the expression of
renormalized viscosity \cite{McCoWatt,McCo:book,MKV:PR}, and it has
been found it to be between 0.35 to 0.40. In this paper we take $\nu^{*}=0.38$
\cite{MKV:PR}. The correlation function $C(k)$ is related to the
one-dimensional energy spectrum $E(k)$:\begin{equation}
C(k)=\frac{{E(k)}}{4\pi k^{2}},\end{equation}
 and $T_{i}'$s are given in \cite{MKV:PR}\begin{eqnarray*}
T_{1} & = & kp(xy+2z^{3}+2xyz^{2}+x^{2}z),\\
T_{2} & = & -kp(xy+2z^{3}+2xyz^{2}+y^{2}z),\\
T_{3} & = & -kq(xz-2xy^{2}z-y^{2}z),\end{eqnarray*}
where $x,y,z$ are cosines defined as \begin{equation}
\mathbf{p}\cdot\mathbf{q}=-pqx,\hspace{1cm}\mathbf{q}\cdot\mathbf{k}=qky,\hspace{1cm}\mathbf{p}\cdot\mathbf{k}=pkz.\end{equation}

The field-theoretic method mentioned above has certain similarities
with the calculations based on the eddy-damped quasi-normal Markovian
approximation (EDQNM). Both these methods use quasi-normal approximation,
and eddy or renormalized viscosity. 

In the subsequent subsections we will use the above formalism to compute
energy fluxes using energy spectrum obtained from a model and direct
numerical simulation. We also estimate the extent of the bottleneck
effect using energy transfer ideas.

\subsection{Bottleneck Effect in Energy Spectrum}

We compare our theoretical results with numerical simulation. The
simulations have been performed for homogeneous, isotropic turbulence
with stochastic forcing at low wavenumbers. These simulations were
done at $512^{3}$, $1024^{3}$, and $2048^{3}$ grids. Taylor-based
Reynolds numbers for these runs were approximately 240, 400 and 700
respectively. (see Yeung et al. \cite{DonzKRS:Scalar} for details
on simulation.) We multiply the numerical energy spectrum with $k^{5/3}/K_{Ko}$
$(K_{Ko}=1.58)$, then divide the resultant quantity by its maximum
value in the inertial range, and obtain compensated energy spectrum
$\tilde{{E}}(k)$. In the inertial range, $\tilde{{E}}(k)=1$. In
Fig. 1 we plot $\tilde{{E}}(k)$ obtained from direct numerical simulations
(DNS) done on $512^{3}$, $1024^{3}$, and $2048^{3}$ grids.  A hump
appears in all the DNS plots indicating the existence of the bottleneck
effect in numerical simulations. These results are consistent with
earlier numerical results showing the bottleneck effect.

Comparison of the normalized energy spectra for different grid resolutions
reveals that the hump is most dominant for $512^{3}$, and it decreases
as the grid size or Reynolds number is increased, a phenomenon observed
in earlier numerical results as well \cite{Kane:Bottle,Bran:Alpha,Gotoh:Scalar_Bottle,Yeung:JFM96,DonzKRS:Scalar}.
This result indicates that the bottleneck effect decreases with the
increase of inertial range, thus reinforcing our hypothesis that the
bottleneck effect may be due to nonavailability of sufficient range
of wavenumbers to facilitate energy cascade. Please note that we have
quantified the bottleneck effect by the size of the hump in the \emph{normalized
energy spectrum}. In individual energy spectrum the size of the hump
could depend on the energy input rate etc. Also, we observe a hump
at low wavenumbers which is due to the forcing at these scales. The
focus of this paper is on the hump at $k\sim k_{d}$ and we will not
analyze the one at the lowest waveumbers.

Let us compare the above energy spectra with a model energy spectrum
for a turbulent flow \cite{Chan:RG,Pope:book,Pope:Hypervisc_Bottle}\begin{equation}
E(k)=K_{Ko}A(k/k_{f})\Pi^{2/3}k^{-5/3}\exp{(-ck/k_{d})},\label{eq:Ek-all}\end{equation}
where \begin{equation}
A(x)=\frac{{x^{s+5/3}}}{1+x^{s+5/3}}\end{equation}
with forcing wavenumber $k_{f}=2$, $c=0.2$, and $s=4$. Throughout
this paper we take $K_{Ko}=1.58$ \cite{Lesi:book,McCoWatt}. Clearly,
$E(k)\propto k^{s}$ for $k<k_{f}$, $E(k)\propto k^{-5/3}$ for the
intermediate range $(k_{f}<k<k_{d})$, and $E(k)\propto k^{-5/3}\exp(-ck/k_{d})$
for the dissipation range $(k>k_{d})$. The choice of $s=4$ is based
on Batchelor's spectra \cite{Davi:book} for smaller wavenumbers.
There is no hump in the model spectrum because of the choice of its
functional form. Here we compare these spectra with spectra that show
the bottleneck effect in order to see how the latter affects the spectral
energy transport.

Without loss of generality we can take $\Pi=1$. In Fig. 1 we plot
$\tilde{{E}}(k)$, which is given by\begin{equation}
\tilde{{E}}(k)=E(k)k^{5/3}/K_{Ko}=A(k/k_{f})\exp{(-ck/k_{d})}\end{equation}
 As expected, $\tilde{{E}}(k)$ with higher $k_{d}$ produces a larger
inertial range.

In the following subsection we will compute energy flux by substituting
the above energy spectra (DNS and model) in Eq. (\ref{eq:Pik0}) and
compare the results. They provide important clues for the bottleneck
effect.

\subsection{Bottleneck Effect in Energy Flux}

First, we compute the flux $\Pi(k)$ by substituting the model energy
spectrum {[}Eq. (\ref{eq:Ek-all})] in Eq. (\ref{eq:Pik0}) with $\Pi=1$.
We compute the integral $I(k_{0})$ (the bracketed term of Eq. {[}\ref{eq:Pik0}])
for various values of $k_{d}$. When $k_{d}=\infty$ and $A(x)=1$,
the integral $I_{\infty}=0.50$ independent of $k_{0}$, implying
that the flux is independent of $k_{0}$ for the Kolmogorov energy
spectrum $(E(k)=K_{Ko}\Pi^{2/3}k^{-5/3})$. Using $I_{\infty}$ we
find the Kolmogorov constant, $K_{Ko}=1.58$ (this is how $K_{Ko}$was
computed in \cite{MKV:PR}). After this the integral $I(k_{0})$ is
computed using the model spectrum with $s=4$, $c=0.2$, and $k_{d}=100,1000,10000$.
The value of $I(k_{0})$ starts from 0 at $k_{0}=0$, reaches a peak,
and then it decays. 

The energy fluxes at various wavenumbers are\begin{equation}
\Pi(k_{0})=K_{Ko}^{3/2}I(k_{0})\end{equation}
 with $K_{Ko}=1.58$. Fig. 2 contains plots of $\Pi(k_{0})$ vs. $k_{0}$
for different values of $(k_{d},c)$. The maximum values of $\Pi(k_{0})$
for these cases are listed in Table 1. They are all less than 1, but
the difference from the actual value (1) is lower for larger $k_{d}$.
Theoretically $max(\Pi(k_{0}))$ must be 1 because the energy input
at small wavenumber is 1. \emph{The reason for the decrease in $max(\Pi(k_{0}))$
is the lack of modes in the inertial range.} This is where the hump
in the energy spectrum near dissipation wavenumber comes into play. 

After the flux calculation for model spectrum, we compute the flux
integral using $\tilde{{E}}(k)$ obtained from DNS at $512^{3}$,
$1024^{3}$, and $2048^{3}$ grids, and obtain $max(\Pi_{DNS})$.
These values are listed in Table 1. The value of $max(\Pi_{DNS})$
for $2048^{3}$ is very close to unity. Clearly the energy spectra
obtained from numerical simulations provide a better handle on energy
flux as compared to the model energy spectrum {[}Eq. (\ref{eq:Ek-all})].
This is because of the higher level of energy spectrum (hump) near
Kolmogorov's wavenumber in the DNS (see Fig. 1), which makes up for
the loss of large wavenumber modes. The overall effect is that the
energy flux in high-resolution DNS is closer to what is expected in
an idealized situation when $k_{d}\rightarrow\infty$. Thus consistency
with Kolmogorov's theory is achieved. The value $max(\Pi_{DNS})$
for $512^{3}$ is somewhat higher than 1, which may be  to approximations
made in our theoretical calculations. 

The DNS plots of Fig. 2 are the fluxes computed by substituting the
DNS energy spectra in Eq. (\ref{eq:Pik0}). This exercise was done
to examine the effects of the bottleneck correction in the flux. In
Fig. 3 we plot the normalized energy flux computed directly from DNS
data on $512^{3}$, $1024^{3}$, and $2048^{3}$ grids. The two figures
match qualitatively, but not quantitatively because of the assumptions
made in the field-theoretic calculation. The coupling of wavenumber
modes in forced, inertial range, and dissipation range is not yet
fully understood to be able to resolve $\Pi(k)$ completely from theory
\cite{Alex:Imprint,Oliv:Simulation,Ayye,Bras:Local}.

In the present subsection we showed that the bottleneck correction
near the dissipation range helps in the effective transfer of energy
flux. In the next subsection we estimate the extent of the bottleneck
effect  to the mechanism proposed in our paper.

\subsection{Estimation of the Bottleneck Effect}

In this subsection we will attempt to estimate the extent of the bottleneck
effect using semiquantitative arguments. Because of a lack of complete
understanding of the coupling between the forced, inertial, and dissipative
scales, this is the best we can do at present.

Verma et al. \cite{Ayye} and Verma \cite{MKV:PR} computed the shell-to-shell
energy transfer rate from $m$th wavenumber shell to $n$th wavenumber
shell ($T_{n}^{m}$) using \begin{equation}
T_{n}^{m}=K_{Ko}^{3/2}\Pi\left[\int_{k\in s_{n}}dkk{}^{2}\int_{p\in s_{m}}dp\int_{|k-p|}^{k+p}dq\frac{{pq}}{4k}\frac{{T_{1}C(p)C(q)+T_{2}C(q)C(k)+T_{3}C(p)C(k)}}{\nu^{*}(k^{2/3}+p^{2/3}+q^{2/3})}\right],\end{equation}
where $s_{m,n}$ are the wavenumber range for the $m$th and $n$th
shell respectively. The wavenumber space is divided into various shells
logarithmically. In Verma et al. \cite{Ayye} and Verma \cite{MKV:PR}
the $m$th shell is $(2^{m/4}:2^{(m+1)/4})$. 

Verma et al. \cite{Ayye} and Verma \cite{MKV:PR} computed $T_{n}^{m}$
\emph{in the} \emph{inertial range} using a similar procedure as described
in the previous subsection. Kolmogorov's spectrum $k^{-5/3}$ was
assumed through out the wavenumber space. They found that the energy
transfer is maximal to the nearest neighbour, yet significant energy
is transferred to other shells. For example, the energy transfer rates
from $m$th shell the shells $m+1$, $m+2$, and $m+3$ are 18\%,
6.7\%, and 3.6\% respectively. The transfer rate decreases monotonically
for more distant shells. 

Let us imagine a wavenumber sphere of radius $R$ somewhat in the
middle of the inertial range. Using the shell-to-shell energy transfer
rates we can compute the energy transfers from the above wavenumber
sphere to $n$ shells adjacent to the sphere (wavenumber range $[R:R*2^{n/4}])$.
Simple algebra shows that the above quantity is \cite{MKV:PR}\begin{equation}
\frac{{Q_{n}}}{\Pi}=\sum_{m=1}^{n}m\times T_{n}^{m}.\label{eq:qn}\end{equation}
In Table 2 we list $Q_{n}$ for various values of $n$. The Table
shows that 42.2\% of the flux is transferred to the three adjacent
shells. \emph{To transfer 99.1\% energy we need 28 shells in the right
of the sphere.} Therefore, we require large number of wavenumber shells
for effective energy transfer, and the bottleneck effect is expected
if the inertial range is insufficient. In this theory, the bottleneck
effect would disappear when there are sufficient wavenumber shells
to enable the complete energy transfer. 

The energy transfer among the wavenumber shells is antisymmetric,
that is $T_{n}^{m}=-T_{m}^{n}$. If we assume the above mentioned
waveumber sphere to be in the middle of the inertial range, we require
approximately $28\times2=56$ shells for an effective energy transfer.
Hence, the inertial range $(k_{max}/k_{min})$ required must be around
$2^{56/4}\approx10^{4}$. Hence our estimate for the minimum length
of the inertial range for no bottleneck effect is approximately four
decades. The range of inertial range in all the experiments and simulations
discussed in this paper is less than four decades, and the bottleneck
effect is observed in all of them. Hence our theoretical estimate
is consistent with the present experimental and numerical results.
We remark that the above estimate of the required inertial range for
zero bottleneck effect could be an overestimate. A realistic estimate
requires a detailed study of energy transfer among modes in the whole
range: forcing, inertial, and dissipation range. 

After the above estimation of required inertial range to suppress
the bottleneck, we move on to estimate the increase in the energy
spectrum due to the bottleneck effect. Suppose the energy spectrum
$E(k)$ till the dissipation wavenumber is\begin{equation}
E(k)=K_{Ko}\Pi^{2/3}k^{-5/3}(1+e(k)),\end{equation}
where $e(k)=\delta E(k)/E(k)$ is the normalized bottleneck correction.
There is a complex interaction between the wavenumbers in the forcing,
inertial, and dissipation range, which is not yet completely understood.
For time being we estimate the additional energy transfer due to the
bottleneck correction to be of the order of $T_{bottleneck}\sim\Pi\times e(k_{d})$.
Since the energy supplied at the large-scales has to reach the dissipation
scale, and if the number of wavenumber shells to the right of above
mentioned wavenumber sphere is $n$, then\begin{equation}
\frac{{Q_{n}}}{\Pi}+\frac{{T_{bottleneck}}}{\Pi}\approx1.\end{equation}
Therefore,\begin{equation}
e(k_{d})=\frac{{\delta E(k_{d})}}{E_{Kolm}(k_{d})}\approx\left[1-\frac{{Q_{n}}}{\Pi}\right].\label{eq:ekd}\end{equation}
Using Zhou \cite{Zhou:Local} and Verma et al.'s results \cite{Ayye}
that $T_{n}^{m}\approx|n-m|^{-4/3}$ for small $(n-m)$, we estimate\begin{equation}
\frac{{Q_{n}}}{\Pi}\approx\alpha n^{2/3}\label{eq:Qn}\end{equation}
 where $\alpha$ is a positive constant. Assuming that the we have
equal number of wavenumber shells to the left and right of the wavenumber
sphere $R$ discussed above, the ratio of Kolmogorov's wavenumber
and forcing wavenumbers is approximately \begin{equation}
\frac{{k_{d}}}{k_{0}}\sim Re^{3/4}\sim(2^{n/4}\times2^{n/4}),\end{equation}
which yields $n\sim1.5\log_{2}Re$, where $Re$ is the Reynolds number
based on Kolmogorov's scale. Substituting this estimate of $n$ in
Eq. (\ref{eq:ekd}) we obtain\begin{equation}
e(k_{d})\sim1-\alpha(1.5\log_{2}Re)^{2/3},\label{eq:ekd_final}\end{equation}
which is plotted in Fig. 4 for a reference with $\alpha=0.09$. The
three points represent the $(Re,e(k_{d}))$ for three DNS discussed
in the present paper. The choice of $\alpha=0.09$ fits best with
the DNS values, and it is consistent with our $Q_{n}$equation {[}Eq.
(\ref{eq:Qn})]. The numerical values of DNS fit quite well with the
theoretical predictions, however, we need more DNS results for a better
test of our theoretical estimate of the bottleneck correction. Also,
for $\alpha=0.09$, $e(k_{d})\approx0$ for $n\approx37$ and $Re\approx10^{7}$.
These estimates are in reasonable agreement with our earlier estimate
of the length of the inertial range for zero bottleneck effect. Our
prediction of $e(k_{d})$ is proportional to $1-\mbox{const}(\log Re)^{2/3}$,
and it differs from the predictions of earlier theories.

Please note that the above expression for $e(k_{d})$ is only a crude
approximation, and could be an overestimate. To better understand
the bottleneck effect we need to understand the coupling among forcing,
inertial, and dissipation scales, as well as other aspects like intermittency. 

The dynamics at the dissipation rate is quite important in the study
of the bottleneck effect. This is evident from the numerical observations
of Lamorgese et al. \cite{Pope:Hypervisc_Bottle}, Biskamp et al.
\cite{Bisk:Bottle_2D}, and Dobler et al. \cite{Bran:Bottle} who
reported that hyperviscosity enhances the bottleneck effect. Since
the extent of inertial range increases with the introduction of hyperviscosity,
it may appear that the bottleneck effect should decrease in the presence
of hyperviscosity. However, that is not the case. This result is possibly
because of the shorter dissipation range in the presence of hyperviscosity,
and the hump in the energy spectrum near $k_{d}$ could help in the
inertial-range energy transfer as well as in the dissipation of energy.
This is an important question to investigate. So far our focus has
been on the physics of energy transfer in the inertial range. A more
detailed study of energy exchange between wavenumbers in the inertial
and dissipative range is required for a conclusive statement \cite{Alex:Imprint,Bras:Local}.

\section{Conclusions}

To summarize, in this paper we investigated the reason for the bottleneck
effect in turbulence. The energy is supplied at large scales, and
it cascades to smaller scales. Recent numerical and theoretical studies
show that even though most of the energy from a given wavenumber shell
goes to the next wavenumber shell, there is a significant energy transfer
to the distant wavenumber shells. We showed that an effective transfer
of energy flux in the inertial range can take place when there is
approximately four decades of inertial range. If the inertial range
is shorter, a hump is created near Kolmogorov's scale (beginning of
dissipation range) which compensates for the nonexistence of required
inertial range. The bottleneck effect is observed in most of the current
numerical simulations and experiments.

The mechanism proposed in the present paper differs from that of Falkovich
\cite{Falk:Bottleneck} and Yakhot and Zakharov \cite{YakhZakh:Cleb}.
Falkovich \cite{Falk:Bottleneck} argued that the bottleneck effect
is due to the suppression of nonlinear interactions by dissipative
modes, and it is present for all dissipative turbulence systems. Falkovich
assumes essentially a local energy cascade in contrast to both local
and nonlocal transfers in our mechanism. In our picture, the energy
is transferred to the dissipative scales not only from its immediate
neighbouring wavenumber shells, but also from the middle of inertial
range. The energy transfer by Kolmogorov's spectrum requires certain
minimum inertial range. If this range is not present, the energy levels
of the modes near Kolmogorov's scales increase to facilitate the energy
transfer. Note that if full range of inertial range is present, the
last wavenumber shells in the inertial range would transfer only a
small fraction of energy flux, and there is no bottleneck effect.
Our theory suggests that the bottleneck effect will disappear if the
inertial range is more than approximately four decades. Yakhot and
Zakharov \cite{YakhZakh:Cleb} and She and Jackson \cite{She:Bottle}
obtained $k^{-1}$ bottleneck correction. Our model purely based on
energy flux differs from these theories as well. Quantitatively, our
prediction for the bottleneck correction $e(k_{d})$ is proportional
to $1-\mbox{const}(\log Re)^{2/3}$, and it differs from the predictions
of earlier theories.

Traditional shell models of turbulence assume local energy transfers,
and have a large inertial range ($2^{15-20}\sim10^{5-10}$). The bottleneck
effect is generally not observed in the shell models. However Biferale
and kerr \cite{Bife:ShellBottle} report the bottleneck effect in
a shell-model ($n=15$) based on Kerr-siggia model. So shell models
with small inertial range could show bottleneck effect, but the bottleneck
effect in shell model is in the spirit of Falkovich's mechanism; there
is not enough dissipative scale to dissipate the cascaded energy. 

The ``real'' turbulence however involves local as well as nonlocal
energy transfers that are not simulated in local shell models of turbulence.
The recent nonlocal shell models \cite{Step:NonlocalShell} attempt
to model these features of turbulence, and it will be interesting
to investigate bottleneck effect in the nonlocal shell models. We
remark that the field-theoretic calculation presented in this paper
is more fundamental than the shell model, and some of it's features
are same as the shell model. Still it is instructive to independently
investigate bottleneck effect using nonlocal shell model.

Many important and unresolved issues are involved in the study of
the bottleneck effect. We need to fully understand the nonlinear coupling
between the forcing, inertial, and dissipative range (see Alexakis
et al. \cite{Alex:Imprint}, Debliquy et al. \cite{Oliv:Simulation},
Verma et al. \cite{Ayye}, Brasseur and Wei \cite{Bras:Local} for
some of the recent attempts). The vortex interactions, intermittency
etc. also come up in the study of the bottleneck effect, and we need
to understand them better as well.

The energy transfer in the turbulence of passive scalar and magnetohydrodynamics
follows similar patterns as in fluid turbulence. The energy transfer
is forward and local, yet significant range of inertial-range is required
for effective energy transfer \cite{Ayye:MHD}. Hence we expect the
bottleneck effect to be present in these systems as well. These projections
are consistent with a strong bottleneck effect observed in numerical
simulations of Watanabe and Gotoh \cite{Gotoh:Scalar_Bottle} and
Yeung et al. \cite{DonzKRS:Scalar,Yeung:JFM96} for passive-scalar
turbulence, and those of Haugen et al. \cite{Bran:Flux} for MHD turbulence.
The bottleneck effect has been observed in electron-magnetohydrodynamic
(EMHD) turbulence, and two-dimensional turbulence (see Biskamp et
al. \cite{Bisk:Bottle_2D} and references therein), but its cause
is possibly more complex. It has been observed that the bottleneck
effect along the transverse and longitudinal directions are different
\cite{Veer:Bottle}; this result still lacks satisfactory explanation.
Future developments in theoretical turbulence will possibly resolve
some of these issues.

\begin{acknowledgments}
We gratefully acknowledge useful discussions with K. R. Sreenivasan
and P. K. Yeung, and hospitality of International Center for Theoretical
Physics (ICTP), Trieste, during our visit in Summer 2005 when part
of this work was done. We thank P. K. Yeung for sharing the numerical
data with us. We also thank one of the referees for very useful suggestions
and comments. The simulations used in this work were done at the San
Diego Supercomputer Center and the National Energy Research Scientific
Computing Center. 
\end{acknowledgments}
\bibliographystyle{unsrt}
%\bibliography{/home/mkv/res/bib/res}

\pagebreak

\begin{table}

\caption{The maximum values of energy fluxes for the model energy spectra
{[}Eq. (\ref{eq:Ek-all})] with $c=0.2$ and $k_{d}=100,1000,10000$,
and for energy spectra obtained from numerical simulations on $512^{3},1024^{3}$
and $2048^{3}$ grids. For Kolmogorov's spectrum $max(\Pi(k_{0}))=1$.}

\begin{tabular}{|c|c|c|}
\hline 
&
$k_{d}$&
$max(\Pi(k_{0}))$\tabularnewline
\hline
\hline 
kd100&
100&
0.84\tabularnewline
\hline 
kd1000&
1000&
0.94\tabularnewline
\hline 
kd10000&
10000&
0.96\tabularnewline
\hline 
DNS512&
-&
1.14\tabularnewline
\hline 
DNS1024&
-&
1.09\tabularnewline
\hline 
DNS2048&
-&
1.02\tabularnewline
\hline
\end{tabular}
\end{table}

\begin{table}

\caption{The energy transfer rates from a wavenumber sphere in the inertial
range to $n$ shells adjacent to the sphere $(Q_{n}/\Pi=\sum_{m=1}^{n}m*T_{n}^{m})$
for various $n$'s. }

\begin{tabular}{|c|c|c|c|c|c|c|c|c|}
\hline 
$n$&
1&
2&
3&
8&
13&
28&
32&
48\tabularnewline
\hline
\hline 
$2^{n/4}$&
$2^{1/4}$&
$2^{1/2}$&
$2^{3/4}$ &
4&
95&
128&
256&
4098\tabularnewline
\hline 
$Q_{n}/\Pi$&
0.18&
0.32&
0.42&
0.74&
0.88&
0.99&
0.99&
$\sim1$\tabularnewline
\hline
\end{tabular}
\end{table}

\textbf{\pagebreak}

\begin{center}\textbf{Figures}\par\end{center}

\begin{figure}
\includegraphics[scale=0.5]{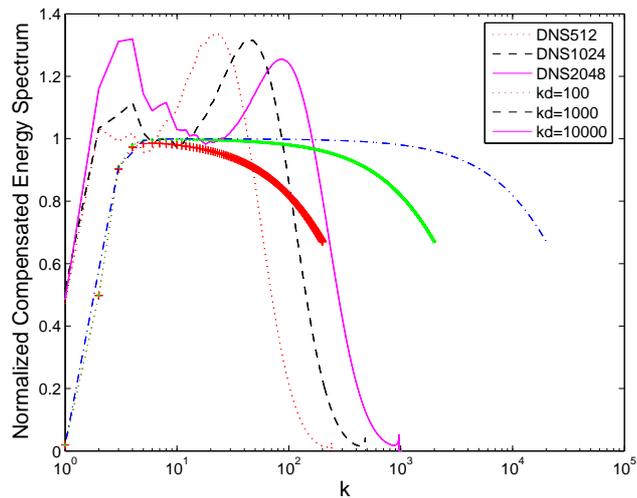}

\caption{\label{Fig:Ek} The normalized compensated energy spectra $\tilde{{E}}(k)=E(k)k^{5/3}/K_{Ko}$
vs. $k$ for a model energy spectrum {[}Eq. (\ref{eq:Ek-all})] with
$c=0.2$ and $k_{d}=100,1000,10000$, and from numerical simulations
on $512^{3},1024^{3}$ and $2048^{3}$ grids at steady state. We take
the energy flux $\Pi=1$ in the inertial range, so that $\tilde{{E}}(k)=1$
in the inertial range.}
\end{figure}

\begin{figure}
\includegraphics[scale=0.5]{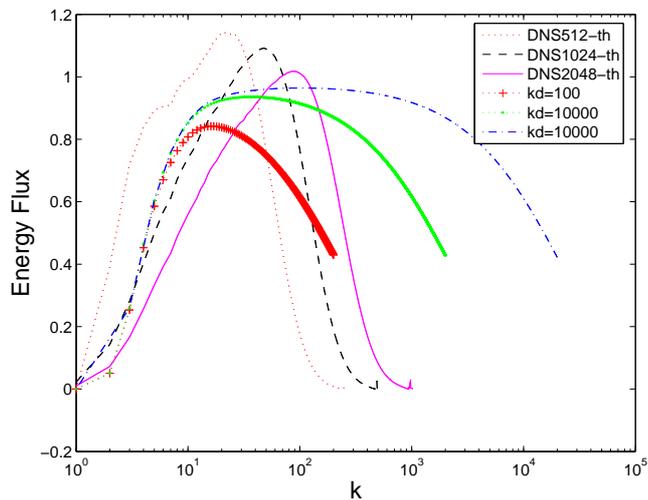}

\caption{\label{Fig:flux} The computed flux $\Pi(k_{0})$ using Eq. (\ref{eq:Pik0})
for a model energy spectrum {[}Eq. (\ref{eq:Ek-all})] with $c=0.2$
and $k_{d}=100,1000,10000$, and from numerical simulations on $512^{3},1024^{3}$
and $2048^{3}$ grids at steady state.}
\end{figure}

\begin{figure}
\includegraphics[scale=0.5]{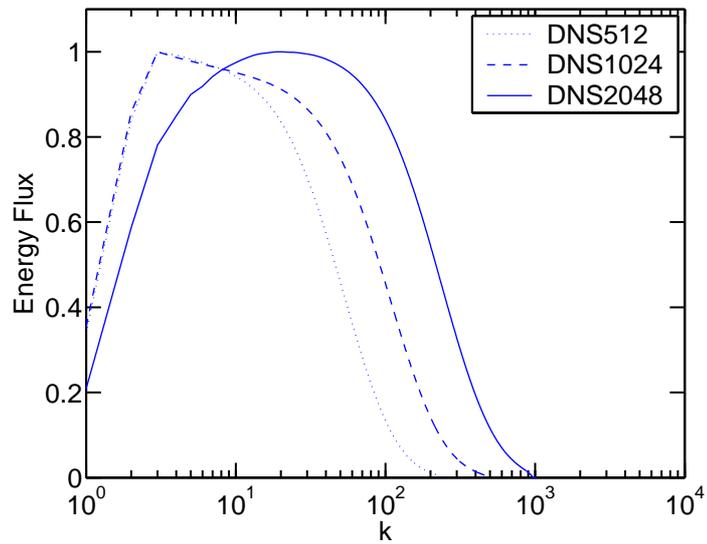}

\caption{\label{Fig:DNSflux} The normalized energy flux computed directly
from DNS on $512^{3},1024^{3}$ and $2048^{3}$ grids under steady
state. }
\end{figure}

\begin{figure}
\includegraphics[scale=0.5]{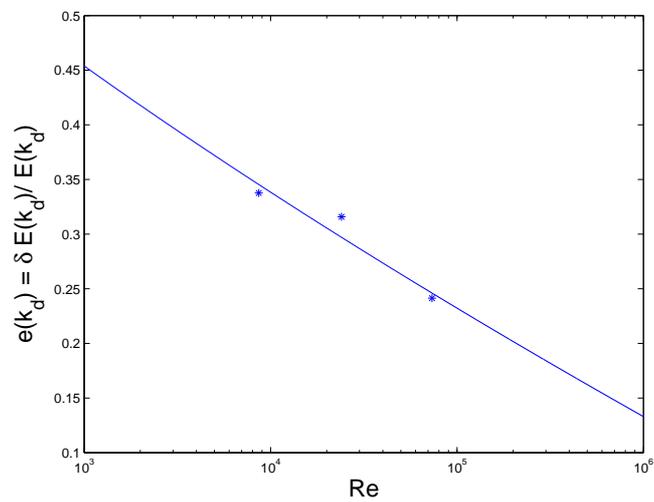}

\caption{\label{Fig:Rebottle} A plot of our estimated normalized bottleneck
correction $e(k_{d})=\delta E(k_{d})/E(k_{d})$ {[}Eq. (\ref{eq:ekd_final})]
as a function of Reynold number $Re$. The three points ($*$) represent
$[Re,e(k_{d})]$ for \emph{DNS} on $512^{3},1024^{3}$ and $2048^{3}$
grids at steady state. }
\end{figure}

\end{document}